\documentclass[twocolumn,aps,10pt,showpacs]{revtex4}
\usepackage{graphicx}
\usepackage{dcolumn}
\usepackage{bm}
\usepackage{color}
\usepackage{hyperref}
\begin{document}
\title{Full 3D Maxwell solver for Inertial Electrostatic Confinement devices}
\author{Victor-Otto de Haan}
\email{victor@bonphysics.nl}
\affiliation{BonPhysics B.V., Laan van Heemstede 38, 3297 AJ Puttershoek, The Netherlands}
\date{November 6, 2026}
\begin{abstract}
Results of a full 3D electromagnetic plasma fusion simulator are presented. It will enable investigations into the influence of structured electromagnetic potentials, anomalous magnetic moments, or charge cluster structures on the fusion yield. The simulator calculates fusion yields similar to those found in real world fusors. In addition, the ion energy and density are accurately simulated. Given that the ion properties arise from a full Maxwell solver there is ample evidence that this part of the simulation is accurate. Concluded with a recommendation for continued research. 
\end{abstract}
\maketitle

\section{Introduction}
Classical “hot” fusion reactions can occur at somewhat lower energy, due to known effects. The overarching principle here is lattice confinement fusion~\cite{Steinetz2020}. In lattice confinement fusion Deuterium in metal hydrides fuses more readily at lower energies than in a plasma. This is thought to be primarily due to electromagnetic screening provided by the lattice potential. In total this screening has demonstrably produced an enhancement of up to a factor of 1000. This effect is also used to increase neutron yield in beam target fusion, where the target is a Deuterated metal~\cite{Raiola2002}. It has been suggested that locally in defects in metals the enhancement due to screening can reach twenty orders of magnitude~\cite{Metzler2024}. This effect may also explain why in some electrostatic confinement fusors a Titanium electrode sometimes results in a higher-than-expected neutron production rate at lower energy. 

These effects are generally produced with a classical fusion fuel, Deuterium, paired with a metal capable of forming a metal hydride or a catalyst. Evidence, where provided, seems to suggest in both metal hydrides and the catalysts the interesting reaction appears to be confined close to the surface of the material. It is known that lattice potentials can screen the repulsive Coulomb interaction, this can be enhanced in defects, but also at surfaces where the dimensionality is reduced from three to two. The latter also increases quantum tunnelling rates~\cite{Metzler2024}. 

Phillips and Ordonez have suggested that when ions bombard a target material a large fraction of kinetic energy is transferred to the target electrons which dissipates as waste, however if electrons are degenerate, they can become effectively transparent as large energies are required to kick them to the next energy level. This enhances fusion yield as it increases the effective centre of mass energy between ion and target~\cite{Phillips2013a,Phillips2013b}. Electron degeneracy of this sort usually only occurs at extreme conditions such as in white dwarf stars. However, in superconductors at low temperatures and sometimes at high pressures electron degeneracy does occur as electrons form cooper pairs. Many of the metal Deuterides discussed previously do become superconducting at low temperature. In addition, Holmlid has suggested that his ultra dense phase of Hydrogen forms a superconductor, based on measurements of a Meissner effect~\cite{Holmlid2015}.

In addition to electron degeneracy, Widom and Larsen posit that localized condensed matter electric fields in metallic hydride surfaces can create electrons with large effective mass~\cite{Widom2005}. Heavier electrons can promote inverse beta decay, giving rise to neutrons, furthermore heavy electrons could, analogous to muons, catalyse fusion events.
Related to this, recent theory and experiment have suggested alternate reaction channels for Deuterium Deuterium fusion near a 0+ threshold resonance. This occurs for very low D-D centre of mass energies. Here the two Deuterium nuclei (spin zero) fuse to a He4 nucleus (also spin zero). The excited He4 nucleus cannot release energy via gamma emission due to angular momentum considerations, so it emits energy through internal pair conversion, resulting in an electron positron pair~\cite{Czerski2016,Dubey2025}. 

It is well known that extremely high electric fields can deform the electrical structure of nuclei, altering the three-dimensional structure of the coulomb barrier. In these cases, fusion yield becomes directivity dependent.  However, nuclear deformation is not necessary, as cited previously, in high power laser driven fusion the large electric field produced by a laser superposed on the nuclear electric fields can provide a lower and thinner coulomb barrier allowing for easier tunnelling.

This is the motivation for producing a particle simulator, which simulates the motion of ions and electrons in an arbitrary potential. 
\section{Physical Description}
Here, a start is made with a simple potential, such as found in an electrostatic fusor to validate the simulator. It is capable of tracking the positions of electrons and ions in an arbitrary geometry and potential. This enables examination of certain potentials for production of exotic ion/charge structures, which may amplify fusion. The fusion rate is calculated using a simple Bosch-Hale model. This fusion cross section is based on empirical observations of the fusion rate against centre of mass energy. Once an exotic structure is identified a separate simulation would be required to calculate the modified fusion cross section. In the next section physical and mathematical background of the particle simulator is described, followed by a code overview and simulation results for various settings.
\subsection{Structure}
The simulator tracks the charges, magnetic moments, positions and velocities of $10^5-10^8$ macro-particles to determine the time, $t$ and position, $\vec{r}$ dependent charge and current density ($\rho(\vec{r},t)$ and $\vec{J}(\vec{r},t)$ respectively) on a three-dimensional discrete Cartesian grid consisting of up to $256 \times 256 \times 256$ cells. The charge and current density are used to solve the inhomogeneous electromagnetic wave equations in vacuum (where $\varepsilon_o$ equals the permittivity and $\mu_o$ the permeability) for the scalar, $\phi$, and vector potentials $\vec{A}$
\[\nabla^2 \phi-\varepsilon_o\mu_o\frac{\partial^2\phi}{\partial t^2}=-\frac{\rho}{\varepsilon_o}\]
\[\nabla^2 \vec{A}-\varepsilon_o\mu_o\frac{\partial^2\vec{A}}{\partial t^2}=-\mu_o\vec{A}\]
by means of a 3D fast Fourier transform in space. The second order time derivative is calculated via finite difference, simply by keeping copies of the potentials two steps backwards in time. The scalar and vector potential are subsequently used to obtain the electric ($\vec{E}_p$) and magnetic ($\vec{B}_p$) fields produced by the macro-particles in the fusor:
\[\vec{E}_p=-\nabla \phi - \frac{\partial \vec{A}}{\partial t}\]
\[\vec{B}_p=-\nabla \times \vec{A}\]
Computationally, the spatial and time derivatives are solved simply with finite difference. At this point an externally applied electric ($\vec{E}_e(\vec{r},t)$) and/or magnetic ($\vec{B}_e(\vec{r},t)$) fields can be added to the field produced by the macro-particles to obtain the total fields
\[\vec{E}_{tot}=\vec{E}_p+\vec{E}_e(\vec{r},t)\]
\[\vec{B}_{tot}=\vec{B}_p+\vec{B}_e(\vec{r},t)\]
One must make sure that the applied fields are solutions to the wave equation applied to the complete geometry of anode and cathode. For most simulations $\vec{B}_e(\vec{r},t)=0$ and $\vec{E}_e(\vec{r},t)=\vec{E}_e(\vec{r})$ is a typical fusor potential where there exists a central spherical grid at some negative potential, while the potential at the edge is set to zero. The 3D fast Fourier transform solves the potentials in a 3D cube and the potentials are set to 0 on the sides of the cube. The cathode and chamber surfaces are defined by the added external potential. As long as charge build-up on the surfaces is relatively small with respect to the space charge this is sufficient for a useful simulation. Adapting the potentials on the conductors so that the electrical potentials are 0 and the vector potentials are parallel to the surface did not significantly influence the simulation results.

Once the total electromagnetic field is determined, the force acting on each macro-particle becomes
\[
\vec{F}=q(\vec{E}+\vec{v}\times\vec{B})
\]
where $q$ is the charge of the macro-particle and $\vec{v}$ its velocity. This force is used to calculated the acceleration and the updated positions and velocities of each macro-particle by means of Euler integration. This can be done classically or semi-relativistically. Larmor precession of the macro-particles magnetic moment can also be calculated.  A Boris-type of integration was also tested (relativistic/non relativistic Boris pusher), but the choice of integrator did not have a noticeable influence on the fusion rates. This is mainly due to the fact that there is no external magnetic field applied. When an external magnetic fields is introduce the role of the integrator must be investigated also.

With the new velocities the energy of each macro-particle is determined and the $D(d,n)^3He$ fusion cross section in a neutral room temperature Deuterium background gas is determined, using an empirical fit due to Bosch and Hale~\cite{Bosch1992}:
\[
\sigma(\epsilon)=\frac{1}{\epsilon}\frac{A_1+\epsilon(A_2+\epsilon(A_3+\epsilon(A_4+\epsilon A_5)))}{1+\epsilon(B_1+\epsilon(B_2+\epsilon(B_3+\epsilon B_4)))}e^{-\sqrt{\epsilon_{G}/\epsilon}}
\]
with $\epsilon$ equals the centre of mass energy in keV and $A_i$,$B_i$ and $\epsilon_G$ the empirical fit parameters. The other D-D fusion reaction,  $D(d,p)^3T$, is left out of consideration here. The cross section can be used to calculate the fusion probability, $fp$ for the given time step, $ \Delta t$ according to a the weighted sum over all macro-particles:
\[
fp=\sum w_i v_i \sigma(\epsilon_i) \rho_o  \Delta t  
\]
where $w_i$ equals the number of ions attributed to the macro-particle $i$, $v_i$  its velocity in the laboratory frame, $\epsilon_i$ half the kinetic energy of its constituent ion (under the approximation that the velocity of the neutral gas molecules can be neglected) and $\rho_o=2p/k_BT$ the neutral gas atom number density with $p$ gas pressure, $T$ temperature, $k_B$ Boltzman constant and the factor two to account for molecular Deuterium.

In a similar way the elastic and inelastic scattering cross section of each particle is calculated. They are used to model the cool down of the plasma, simply by subtracting a small fraction of the particles total energy related to the total cross section seen by the particle.

As a final step the simulator checks for very slow particles, particles that are effectively so cold that they could recombine and neutralize. This is done probabilistically, using a two-dimensional logistic density function. The colder a particle is and the longer it has been cold the higher the probability is that it will neutralize:
\[
E \geq E_0 \ \ \ : \ \  \ p(E,C)= 0
\]
\[
E < E_0 \ \ \ : \ \ \  p(E,C)=\left(1+e^{\frac{E-E_o}{\Delta E}}\right)^{-1}\left(1+e^{\frac{C-T}{\Delta T}}\right)^{-1}
\]
where $E$ equals the particle kinetic energy, $C$ the number of steps the particle has been “cold” (termed coldness), $E_0$ and $T$ the transition energy and coldness respectively and $\Delta E$ and $\Delta T$ characterising the width of the distribution. When a particle is neutralized, it is removed from the simulation and a new particle is spawned at a random location on the grid. This neutralization/reinjection rule is introduced to obtain a realistic discharge mechanism. In principle one should used full recombination / ionization cross sections of the available molecular and ion species. However, such calculations are very sensitive to the composition of the gas and impurities present. Hence, here for a more phenomenological approach is chosen. A bases for the approach is that for a stable discharge each particle that is neutralized must be replaced by a particle that gets ionized at a random position inside grid. This mechanism together with the predetermined ionisation factor of the gas determines the dynamics of the charge/current distributions. Here we used $E_0 = 1$~keV;  $\Delta E = 1$~keV; $T = 100$; $\Delta T = 100$. Hence it only influence the particles with a kinetic energy below 1~keV where the fusion rate is negligible. Note that the chosen probability with an energy below the threshold only slightly depends on the energy and coldness parameter. In future this distribution can be tuned to fine tune the simulation results depending on gas composition and impurities present.

After completing all of these steps, the simulator starts again at step one, determining the charge and current density, using the updated particle positions and velocities, to determine the electromagnetic potentials. The time step ($\Delta t$) has to be smaller than the time it takes for light to cross a single cell (size $\Delta x$). Hence, $\Delta t<\Delta x/c$ where $c=(\varepsilon_o\mu_o)^{-1/2}$ is the velocity of light. Here this is about one picoseconds.

It is important to note that the positions and velocities of each macro-particle is a continuous variable (to within double precision), however the charge and current density are calculated on the $M \times M \times M$ grid which in terms of resolution is much coarser than double or single precision. This is computationally much cheaper than calculating the field at the position of the $m^{th}$ particle produced by $N-1$ particles, $N$ times for each time step.
\subsection{Macro-particles}
A macro-particle represents a cluster of real micro-particles such as electrons or ions. Computationally the macro-particle is a structure containing the following information:
\begin{itemize}
\item Mass of a single micro-particle
\item Charge of a single micro-particle
\item Averaged magnetic moment (vector) of all the micro-particles 
\item $x, y$ and $z$ positions and velocities of the cluster
\item Number of micro-particles contained within the cluster (also called weight)
\end{itemize}
At the time of this report a macro-particle is always purely electrons or purely Deuterium ions, though in principle it is possible to use mixed clusters. The simulation is conducted such that the number of positive and negative macro-particles are equal, making the plasma as a whole neutral. The magnetic moments can be turned off to save major computational time as they do not have a significant effect on the simulation results when there is no magnetic field added (i.e. $\vec{B}_e=\vec{0}$).
\subsection{Basic Code Overview}
The simulation executes the following basic steps:
\begin{itemize}
\item[(1)] 	All fields and potentials are set to zero.
\item[(2)] 	The particle positions are matched to their nearest grid cell and the charge and current of each macro-particle is distributed over the eight neighbouring  corner cells.
\item[(3)] 	Solve the scalar and vector potentials using the wave equation and an FFT.
\item[(4)] 	Using these potentials the effective electromagnetic field is calculated.
\item[(5)] 	Update the particles positions and velocities using the effective electromagnetic field (here also the external applied field at the position of the particle is superimposed analytically).
\item[(6)] 	Next the fusion cross section and probability are calculated using Bosch-Hale. In addition, the scattering probability is calculated and a stochastic energy loss to each macro-particle is applied.
\item[(7)] 	The final re-injector function checks the energy of each macro-particle. If this energy is below a certain threshold a “cold-time” counter starts. Based on the cold-time and the energy a probability is calculated that the particle neutralizes. A random number is drawn to determine whether or not that particle neutralizes. If the particle neutralizes a new particle is spawned at a random location in the chamber with zero kinetic energy (it will begin to accelerate in the applied field).
\item[(8)] 	Repeat 1-7.
\item[(9)] 	Occasionally the particles are sorted according to their positions. This speeds up the calculation process.
\end{itemize}
All calculations are done on a GPU which has thousands of processing units. Since each particle or grid cell can be treated independently, loops over particles and grid cells are made to be parallel and run at the same time on different processing units. Sometimes two processing units need to manipulate the same grid cell, which can obviously lead to computational instability. To this end we use “atomic” operations. In this case the computation is more careful to make sure that at no time two processing units try to modify the same variable. The GPU has hardware support for atomic operations up to float precession. Double precision for atomic operations is not supported at the hardware level, but it can be simulated. For this reason, in all scripts that require atomic operations we convert our double variables to float and then back to double.
\subsection{Influence simulation parameters}
All simulations based on time step size, grid size and number of particles and particle size are prone to stability issues based on the exact choices of these parameters. It is a trade-off between accuracy of the results and calculation time required to obtain the results with available resources and the objective of the calculations. Here, the basis for the choices of the used values are based on the influence on the fusion yield. It was estimated that further enhancement of the accuracy of the simulation does not change the order of magnitude for the fusion rate. Precise criteria for the simulation parameters (for instance time step size, grid size and number of particles and particle size) and their influence on the accuracy of the simulation needs to be developed.
\subsection{Visualization}
All data produced by the simulation can be visualised. A completed 3D geometry can be inspected (see figure~\ref{figVis}A) or a section can be chosen (see figure~\ref{figVis}B). It can be complemented with the location of the particles in the gird (see figures~\ref{figVis}C and D) and the velocity of the particles can be visualised at each time step by changing their position for increasing time starting from the position as the inspected time step (see figures~\ref{figVel}A to C). This enable the inspection of the particle dynamics at each time step. 
\begin{figure}[hbtp]
\begin{picture}(200,260)\put(10,142){
\put(-20,0){\scalebox{0.5}{\includegraphics{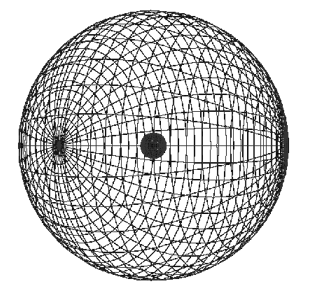}}}
\put(-20,100){(A)}
\put(100,-5){\scalebox{0.5}{\includegraphics{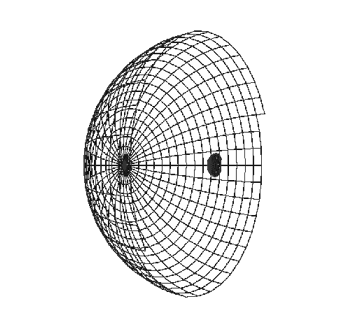}}}
\put(120,100){(B)}
}
\put(10,-15){
\put(-30,0){\scalebox{0.5}{\includegraphics{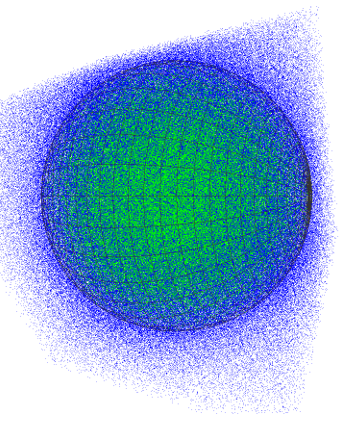}}}
\put(-20,140){(C)}
\put(115,12){\scalebox{0.5}{\includegraphics{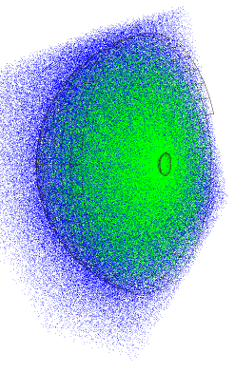}}}
\put(120,140){(D)}
}
\end{picture}
\caption{\label{figVis}Example of visualization of simulation: A completed 3D geometry can be inspected (A) or a section can be chosen (B), complimented with the location of the particles in the gird (C) and (D). Green dots represent Deuterium macro particles; blue dots represent electron macro-particles.}
\end{figure}
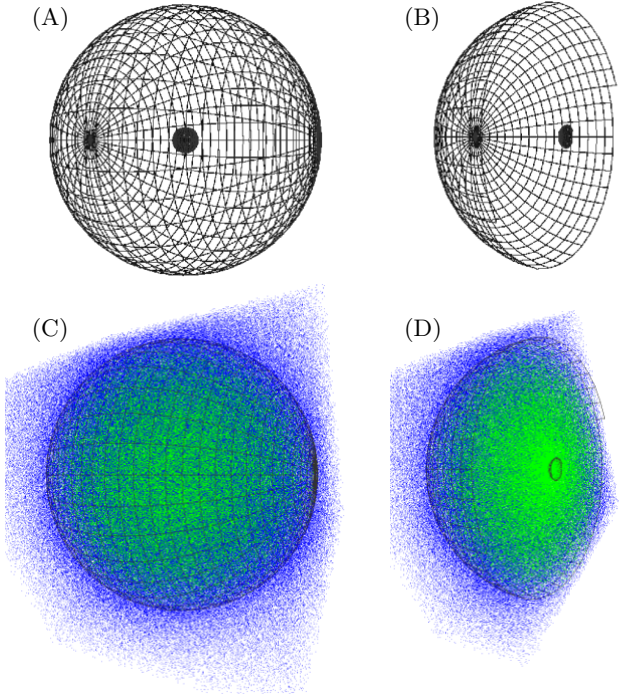
\begin{figure}[hbtp]
\begin{picture}(200,155)
\put(10,-10){
\put(0,0){\scalebox{0.8}{\includegraphics{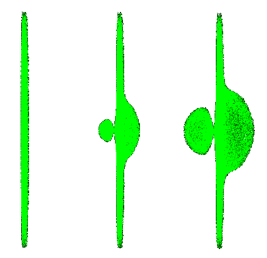}}}
\put(-10,140){(A)}
\put(50,140){(B)}
\put(110,140){(C)}
}
\end{picture}
\caption{\label{figVel}Example of visualization of the velocity of macro-particles in the simulation: (A) side view of section of initial location of the particles at the simulation time step. (B) locations after a certain time, (C) locations at a later time.}
\end{figure}
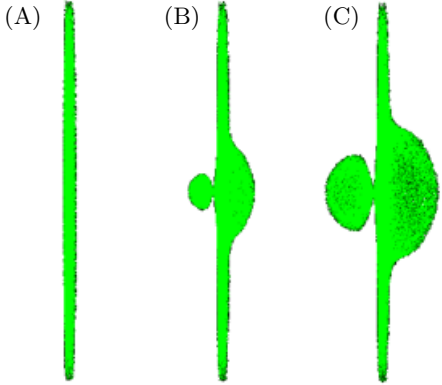
\section{Simulation Results}
\begin{figure}[hbtp]
\begin{picture}(200,150)
\put(10,0){\scalebox{0.5}{\includegraphics{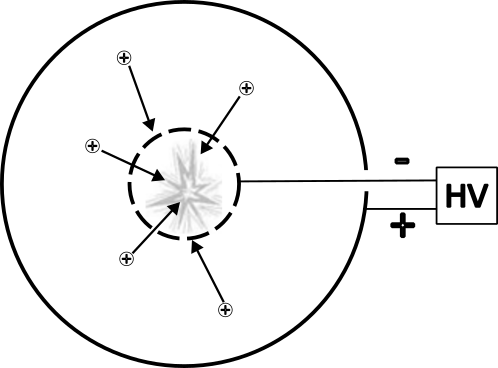}}}
\end{picture}
\caption{\label{fig1} Simple schematic of a Farnsworth Fusor. A spherical chamber is charged with a low pressure Deuterium gas and a high negative voltage applied to the central grid which ionizes a fraction of the gas. Deuterium ions are accelerated to fusion energies and collide with the surrounding unionised gas leading to fusion.}
\end{figure}

To validate the simulator a standard Farnsworth fusor (see figure~\ref{fig1}) with a spherical central grid and a spherical wall is simulated. The fusor is assumed to be loaded with a Deuterium gas (at room temperature and a pressure of 1.38 Pa) and the ionisation fraction is assumed to be small $10^{-7}$ (dilute plasma). Below the charge structure for various applied voltages and grid sizes is examinated. In all cases the fusor has a radius of 5 cm. The radius of the cathode and the applied voltage is varied. $10^7$ macro-particles  are used on a grid of $256 \times 256 \times 256$ cells. The time step is 1~ps and the total simulation time is 50 ns, after which an equilibrium has been reached. Depending on the geometry, a simulation with this number of cells and macro-particles takes between 14 and 24 hours on a NVIDIA Tesla V100-SXM2-16GB~\cite{TeslaV100}.
\subsection{5 mm radius cathode at -100 kV}
\begin{figure}[hbtp]
\begin{picture}(200,170)
\put(-25,-10){\scalebox{0.33}{\includegraphics{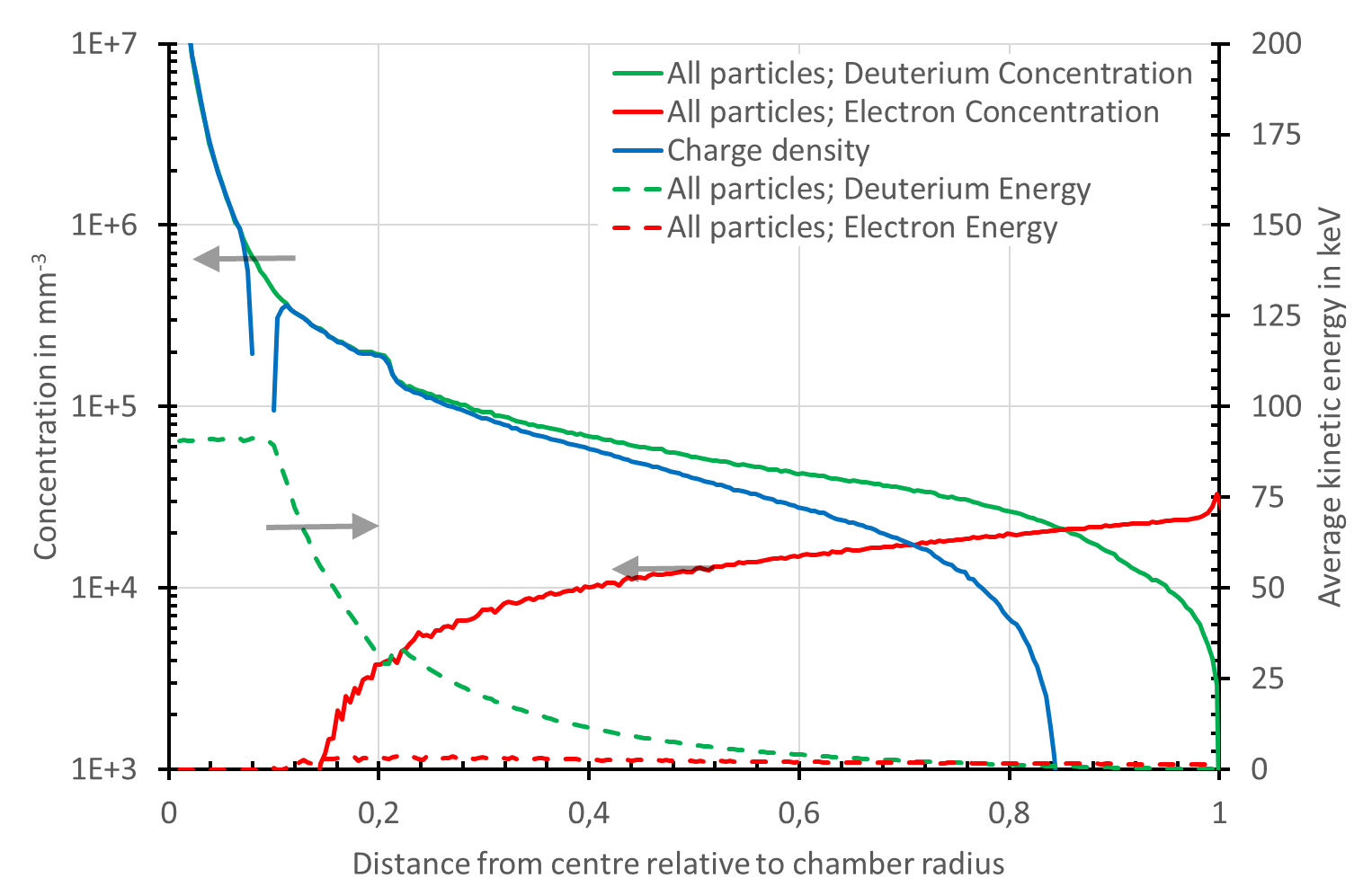}}}
\end{picture}
\caption{\label{fig5mm100kV} Charge and particle energy distribution as function of relative distance to the centre of a fusor with radius of 5 cm and a cathode radius of 5 mm held at -100 kV. The left vertical axis shows the particle concentration (full lines) and the right one shows the average kinetic energy per particle (dashed lines).}
\end{figure}

The negative cathode at the centre has a radius of 5 mm, a thickness of 1 mm and is held at -100 kV. The results ot the simulation are shown in figure~\ref{fig5mm100kV}. Interestingly here the steady state charge density shows a bump at a relative distance of 0,2 (i.e. at 1 cm) from the centre of the fusor (i.e. at a distance of about 5 mm from the cathode). At a relative radius of 0,1 one can see that the charge density becomes 0 (this is inside the cathode) and a the centre the charge density is positive again, where the density is increasing towards the centre. Within the volume that is surrounded by the cathode, the electron concentration is 0: they are always pushed away from the cathode. There, the average energy of the Deuterium ions is constant ($\approx 91$~ keV) as the electric field is 0 the Deuterium ions are not accelerated any more.
\subsection{10 mm radius cathode at -100 kV}
\begin{figure}[hbtp]
\begin{picture}(200,170)
\put(-25,-10){\scalebox{0.33}{\includegraphics{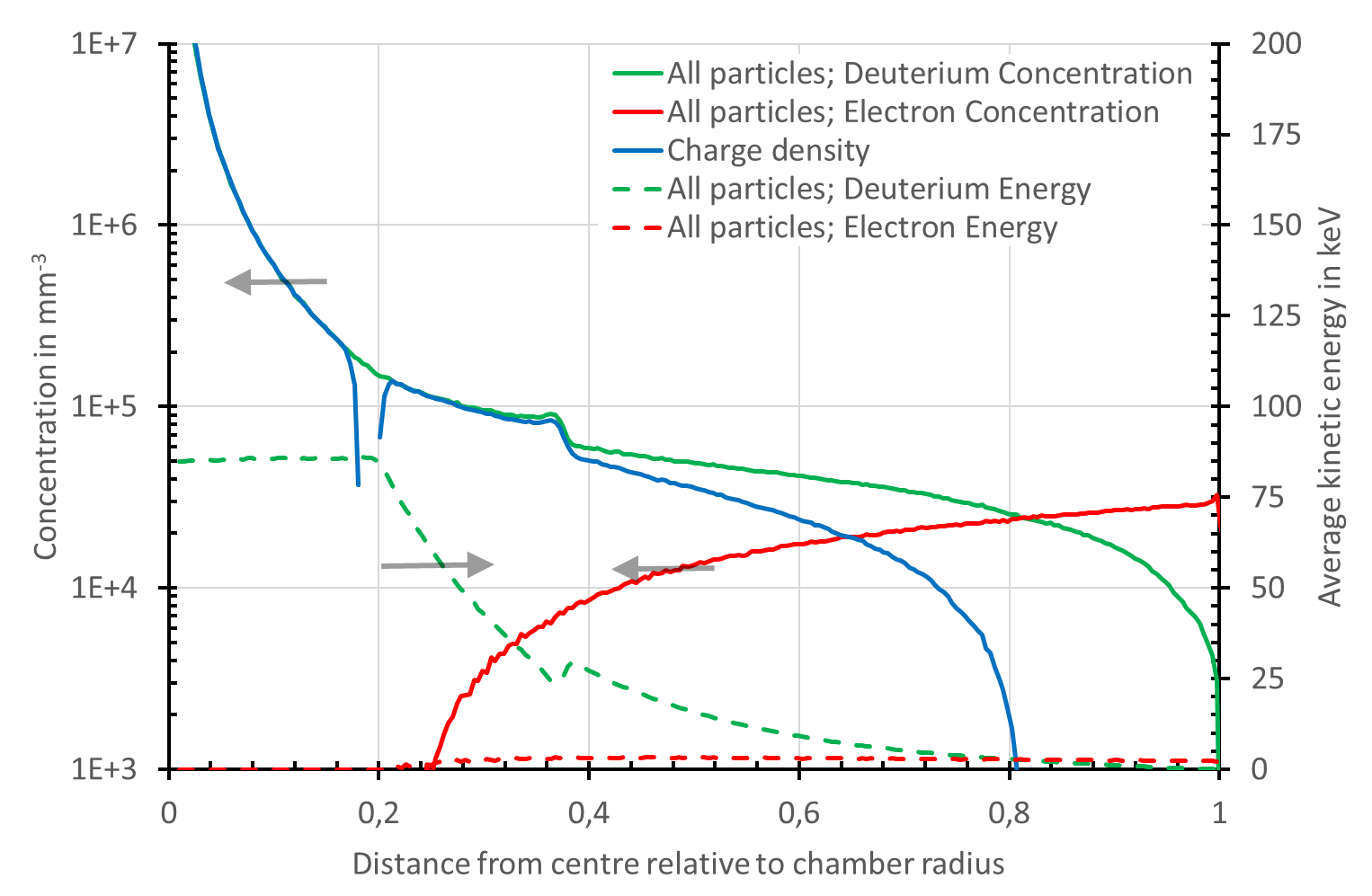}}}
\end{picture}
\caption{\label{fig10mm100kV} Similar to figure~\ref{fig5mm100kV}, except the radius of the cathode is now 10 mm.}
\end{figure}

In this simulation the negative cathode at the centre has a radius of 10 mm, a thickness of 1 mm and is held at -100 kV. The results are shown in figure~\ref{fig10mm100kV}. The bump in the steady state charge density has shifted to at a relative distance of 0,36 (i.e. at 1,8 cm) from the centre of the fusor (i.e. at a distance of 8 mm from the cathode). Within the volume that is surrounded by the cathode, the average energy of the Deuterium ions is again constant ($\approx 85$~keV) and their concentration is approximately a factor $1,4$ larger than before.
\subsection{5 mm radius cathode at -200 kV}
\begin{figure}[hbtp]
\begin{picture}(200,170)
\put(-25,-10){\scalebox{0.33}{\includegraphics{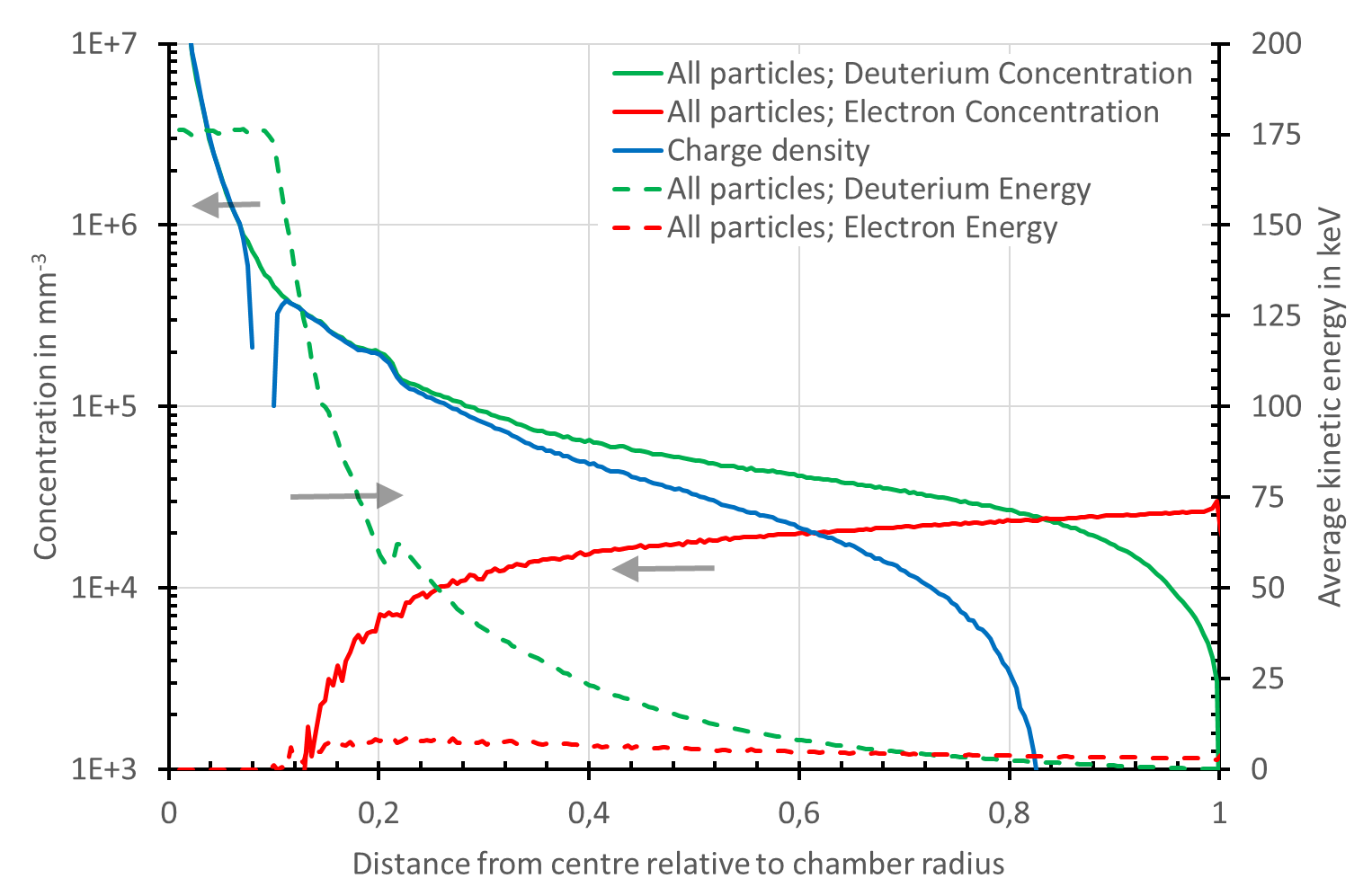}}}
\end{picture}
\caption{\label{fig5mm200kV}  Similar to figure~\ref{fig5mm100kV}, except the cathode voltage is now -200 kV.}
\end{figure}

To investigate whether the charge structure initially seen in the previous sections is the result of a specific electric field strength, in figure~\ref{fig5mm200kV} the results of a simulation with a 5 mm cathode, but with increased field strength due to an applied voltage of -200 kV. The bump returns to approximately the same position as with the lower field strength but the same geometry, giving credence to the idea that the field strength itself is not relevant to the charge structure.  Within the volume that is surrounded by the cathode, the average energy of the Deuterium ions is again constant ($\approx 176$~keV) and their concentration is approximately the same as in the first simulation (factor $1,05$).
\subsection{5 mm radius cathode at -50 kV}
\begin{figure}[hbtp]
\begin{picture}(200,170)
\put(-25,-10){\scalebox{0.33}{\includegraphics{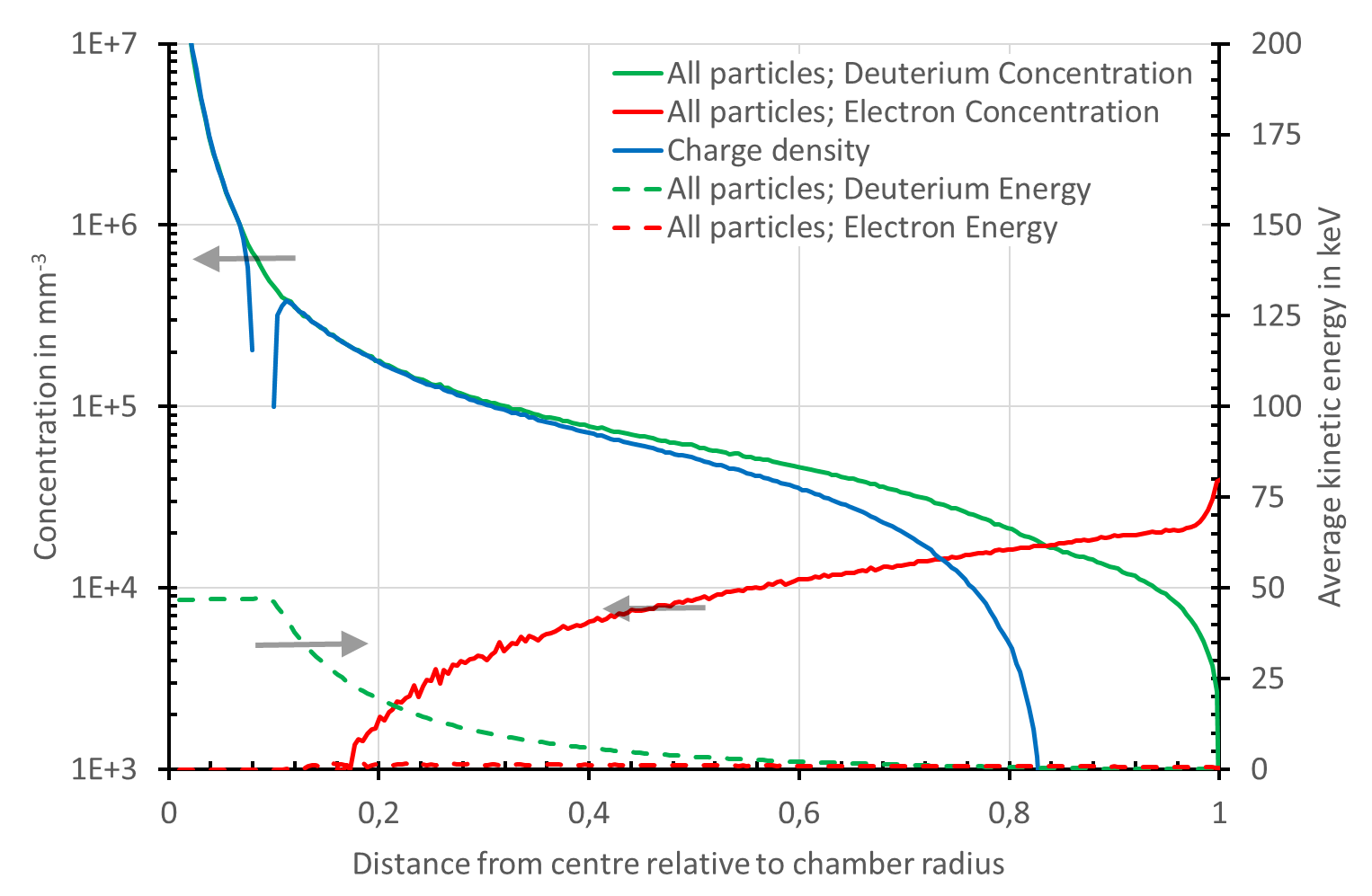}}}
\end{picture}
\caption{\label{fig5mm50kV} Similar to figure~\ref{fig5mm100kV}, except the cathode voltage is now -50 kV.}
\end{figure}

Next, the electric field strength was lowered using an applied voltage of -50~kV. The bump disappears as shown in figure~\ref{fig5mm50kV}.  Within the volume that is surrounded by the cathode, the average energy of the Deuterium ions is again constant ($\approx 47$~keV) and their concentration is approximately the same as in the first simulation (factor $1,05$).
\subsection{Fusion Yield for several geometries}
\begin{figure}[hbtp]
\begin{picture}(200,170)
\put(-25,-10){\scalebox{0.33}{\includegraphics{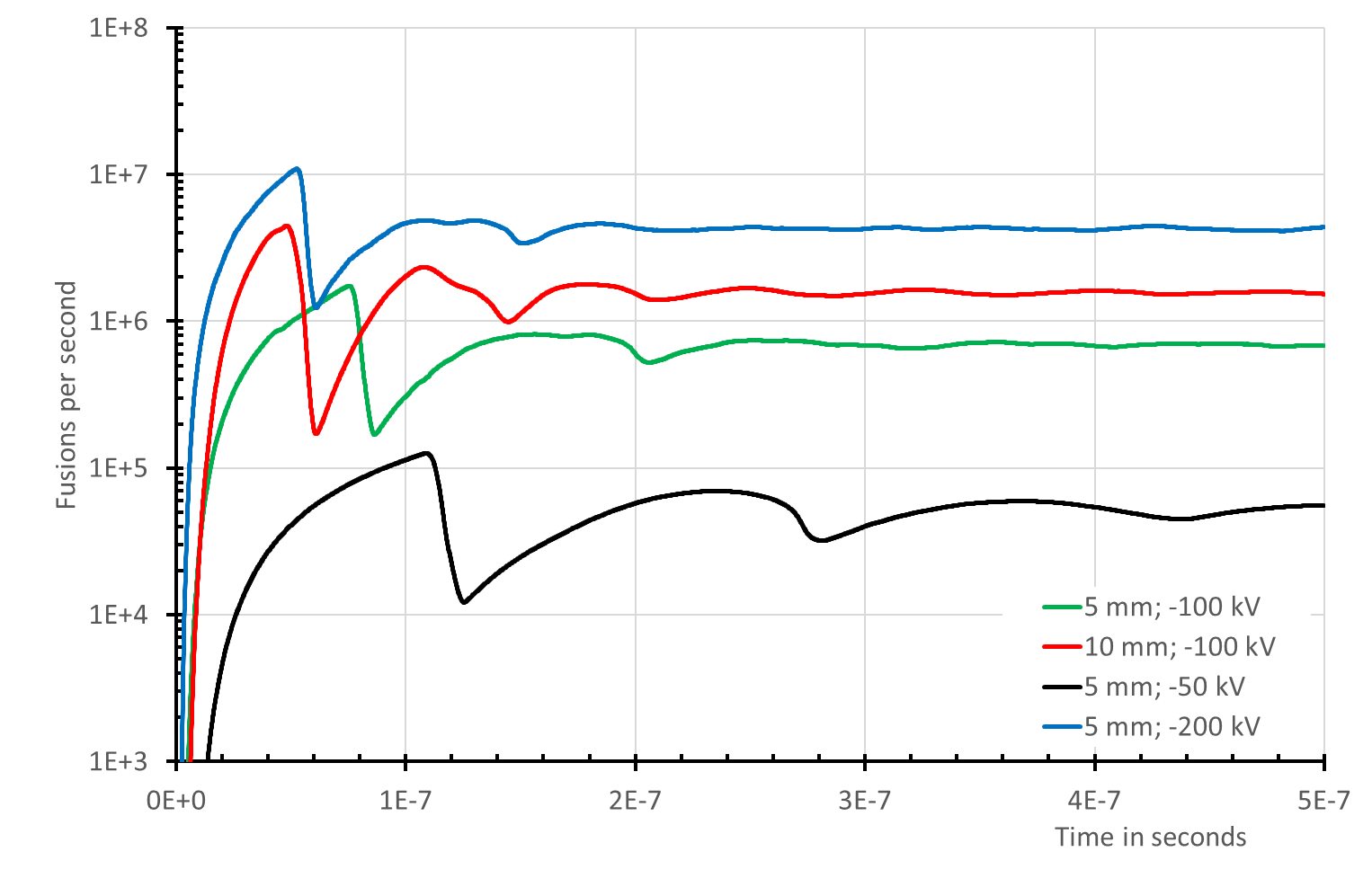}}}
\end{picture}
\caption{\label{figFP} Fusion per second against time steps for the 4 simulations.}
\end{figure}

\begin{figure}[hbtp]
\begin{picture}(200,170)
\put(-25,-10){\scalebox{0.33}{\includegraphics{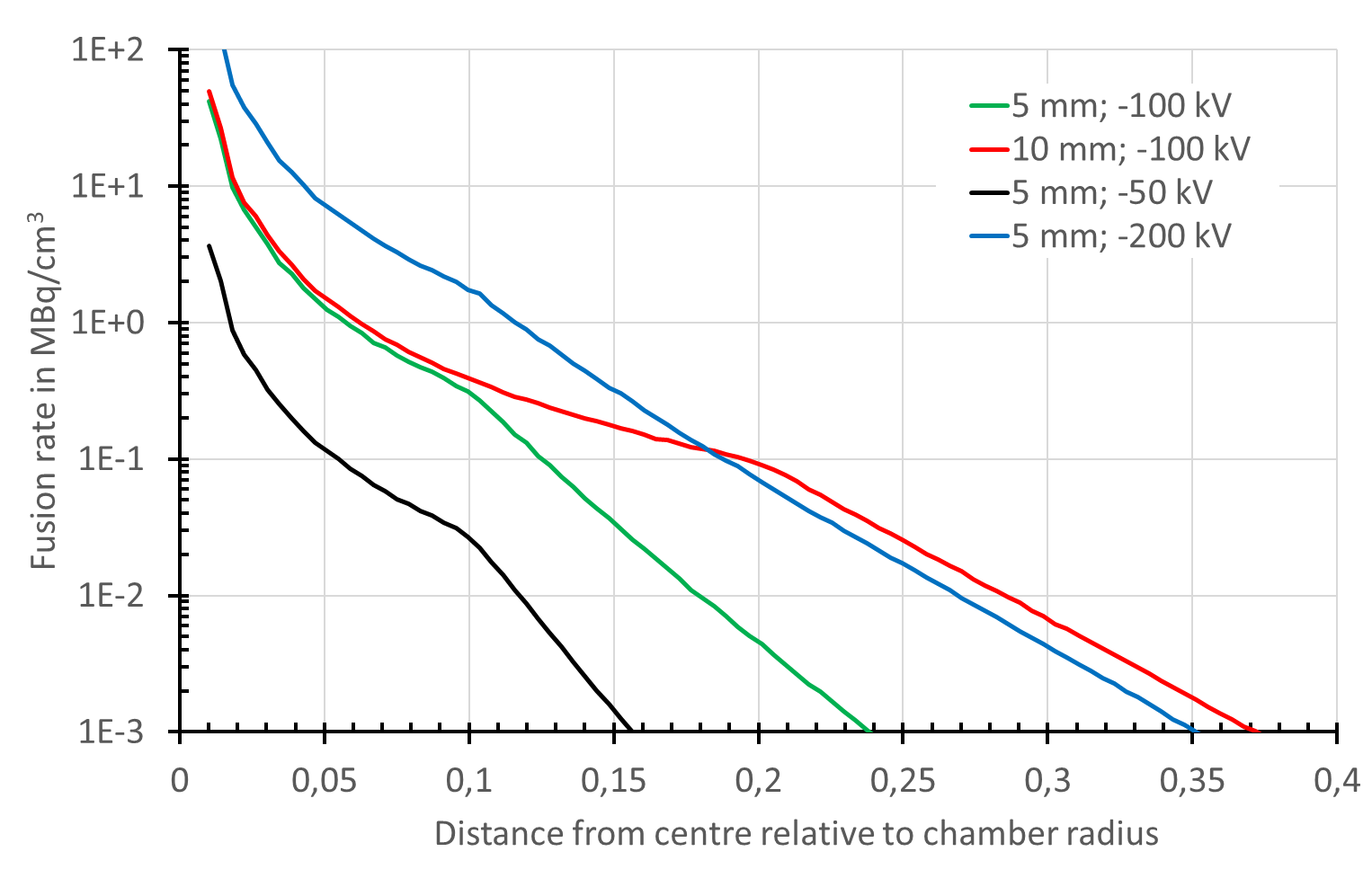}}}
\end{picture}
\caption{\label{figFPdens} Fusions in MBq/cm$^3$ as function of the distance to the centre of the fusor for the 4 simulations.}
\end{figure}

The fusion rate as function of time is shown in figure~\ref{figFP}. After an initial phase, where the simulation starts from an initial distribution of a homogeneous ion distribution with no kinetic energy, the fusion rate stabilizes within 50 ns to a distribution with an energy and concentration distribution as shown it the previous sections. The fusion rates are consistent with high power fusors operated at voltages between 50 and 200 kV~\cite{Wulf2025}. The simulations could overestimate fusion by a small margin, since grid losses are not simulated.
In figure~\ref{figFPdens} the distribution of the fusion rate as function of the distance to the centre of the fusor is shown for the 4 simulations. There are some features in these distributions that require explanation.  

First, it is clear that with increasing field strength (and the same geometry) the fusion rate increases. The fusion rate increases roughly with a factor of 10 when the voltage difference increases with a factor of 2. This is typical for the exponential dependence of the cross section of the D-D fusion on the energy of the centre of mass of the reaction. 

Second, for all simulations there is a discrete change in the slope of the curve at the location of the cathode. This is due to the fact that within the cathode the kinetic energy of the Deuterium ions is constant and does not increase any more when the distance to the centre decreases, while outside the cathode this is still the case. 

Third, the fusion yield inside the cathode does not seem to be dependent on the geometry of the cathode as is evident from the coalescence of the red and green curve for a radius less than that of either cathode. Towards the centre the fusion rate increases to very large values. As the average energy is constant in this region, the increase in fusion rate is due to an increased concentration. This is due to the fact that the fusor accelerates all ions towards the centre of the fusor, to that the ion density is increased, however the number of ions that cross a sphere around the centre is fairly constant so that the total number of fusions in only proportional to the volume. 

Lastly, it is remarkable that the yield outside the cathodes of the fusor with a voltage difference of 200 kV and a cathode radius of 5 mm is smaller than the yield of the fusor with a voltage difference of 100 kV and a cathode radius of 10 mm. The reason for this is that, although the concentration of Deuterium ions is a little lower, the average energy is $15\%$ larger, which increase the fusion chance. The average energy is larger, because the electric field strength is stronger.
\subsection{Deuteron and Electron Average Energy and Distribution}
\begin{figure}[hbtp]
\begin{picture}(200,330)
\put(-25,170){\scalebox{0.33}{\includegraphics{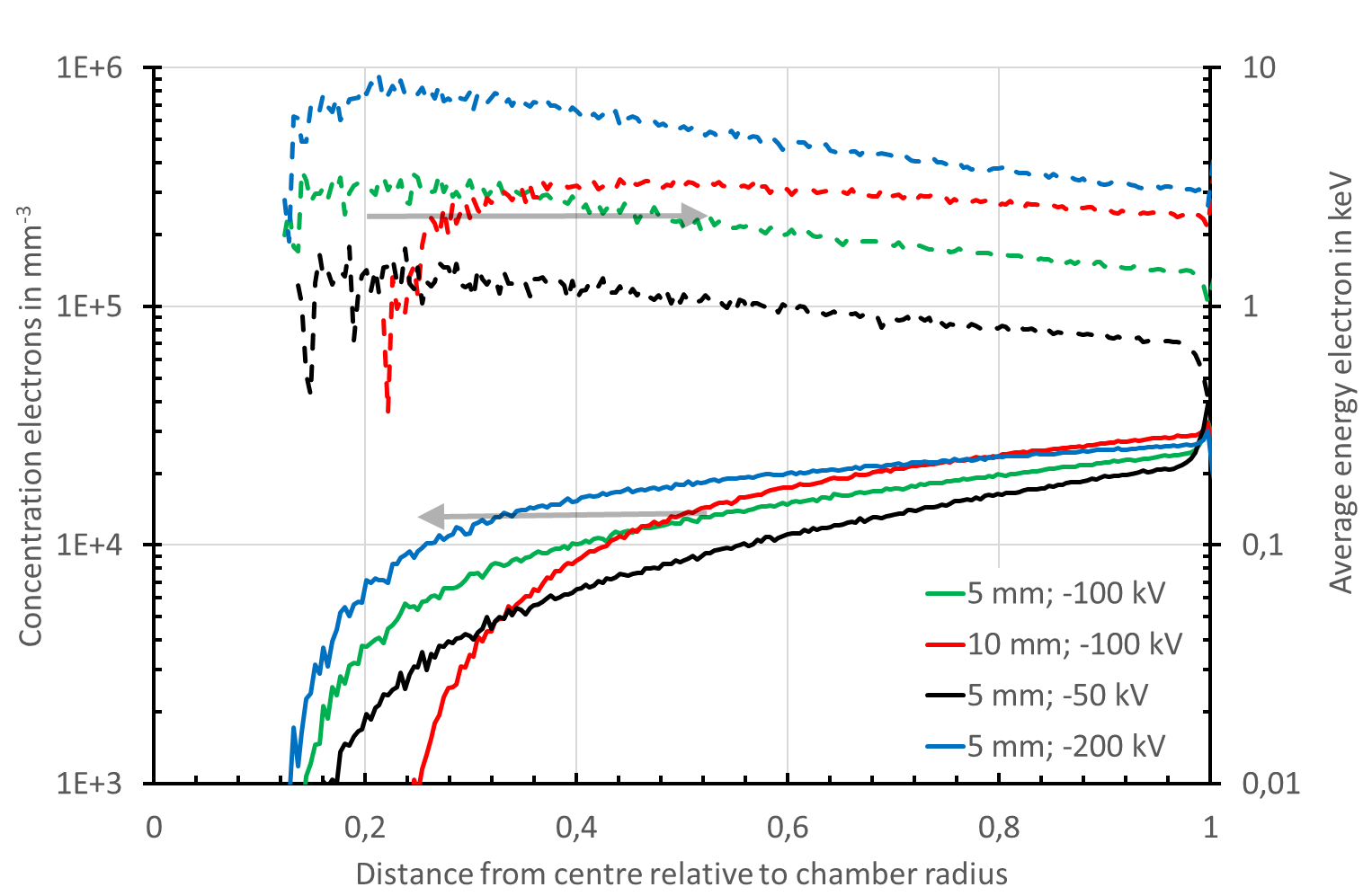}}}
\put(-25,0){\scalebox{0.33}{\includegraphics{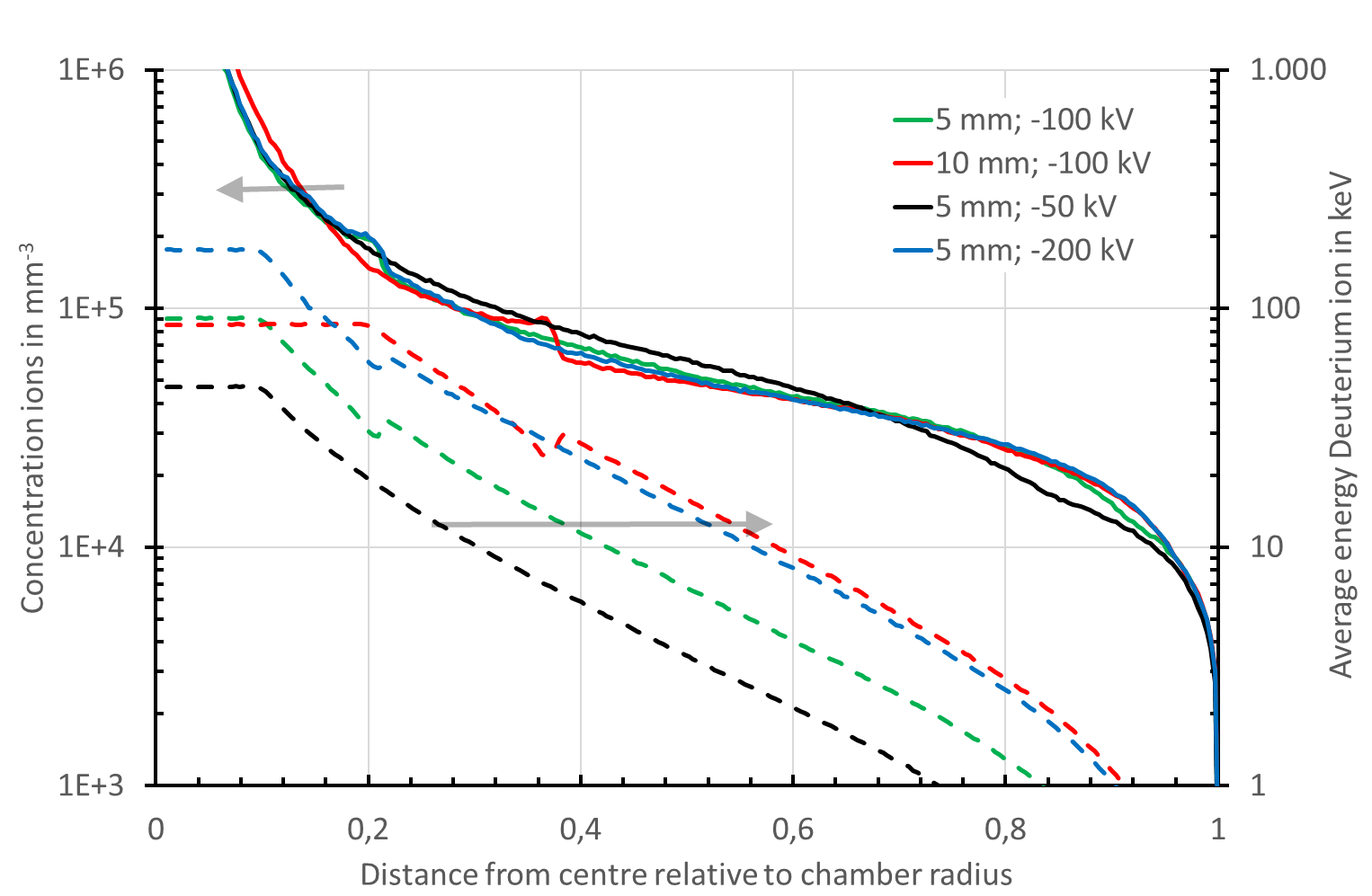}}}
\end{picture}
\caption{\label{figEandC} Particle average energy and concentration distribution as function of relative distance to the centre of a fusor with radius of 5 cm for the 4 simulations: Top: Electrons; Bottom: Deuterium.The left vertical axis shows the particle concentration (full lines) and the right one shows the average kinetic energy per particle (dashed lines).}
\end{figure}

In figure~\ref{figEandC} the deuteron ion and electron kinetic energy as function of distance to the centre is plotted together with their distributions in the fusor.  Electrons tend to move away from the cathode. Their average energy drops when the distance to the centre increases because the electrons that are ionized further away from the centre of the fusor will be able to gain less energy, reducing the electron average energy. This occurs partly due to the fact that electrons that are lost due to capture or escape are spawned randomly through the accessible volume to simulate ionization. In a future expansion one could simulate the ionisation by introduction of seed electrons and ionisation cross sections, so that this part of the simulation becomes more realistic. For now, it is assumed that the average ionisation can be manipulated by changing the gas pressure.
\section{Discussion}
\subsection{Average energy inside cathode}
It is evident from the above that most fusion events occur inside the cathode volume. In these simulations the concentration in this region is mainly determined by the ionisation fraction that was taken, although the geometry has some influence. The main contributing factor is the average kinetic energy of the ions inside this region which is always a little less than what could be obtained with the applied voltage. Partly this is due to the fact that the ions are created throughout the volume of the fusor and partly this is due to the energy exchange of the ions with the background gas. This also limits the maximum applied voltage as the energy loss is higher, the higher the applied voltage (i.e. 3, 9-15 and 24 ~keV for an applied voltage of -50, -100 and -200~kV).

\subsection{Volume of cathode}
When the volume of the region inside the cathode is increased (with the same applied voltage), the fusion rate inside this volume only slightly depends on this volume. This is an indication that the yield of a fusor would increase when this volume is increased. The optimize volume will depend on the average energy loss of the ions inside this volume hence will be smaller for higher applied voltages.

\subsection{Structure in discharge}
\begin{figure*}[hbtp]
\begin{picture}(500,330)
\put(-5,170){\scalebox{0.33}{\includegraphics{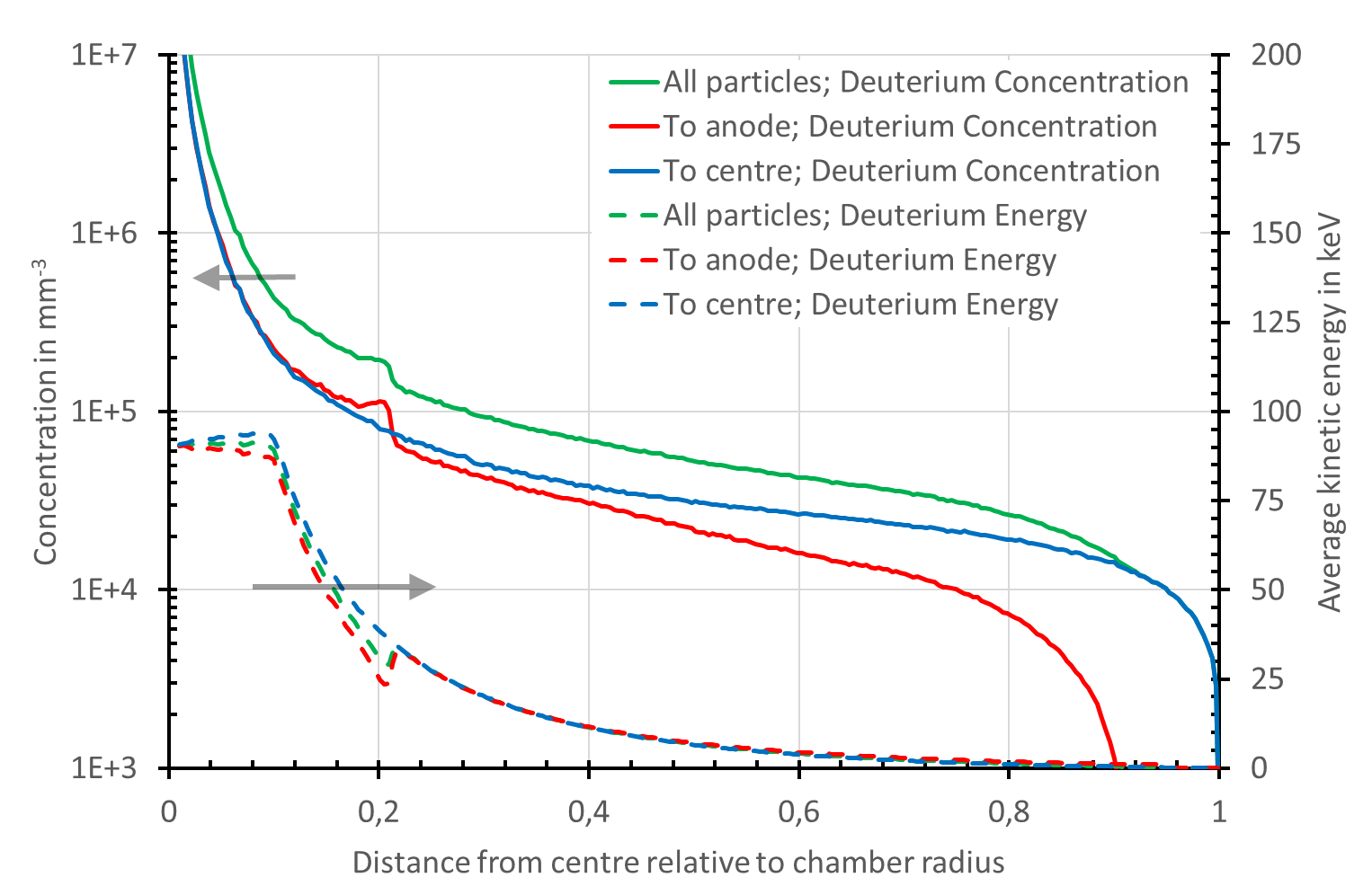}}}
\put(255,170){\scalebox{0.33}{\includegraphics{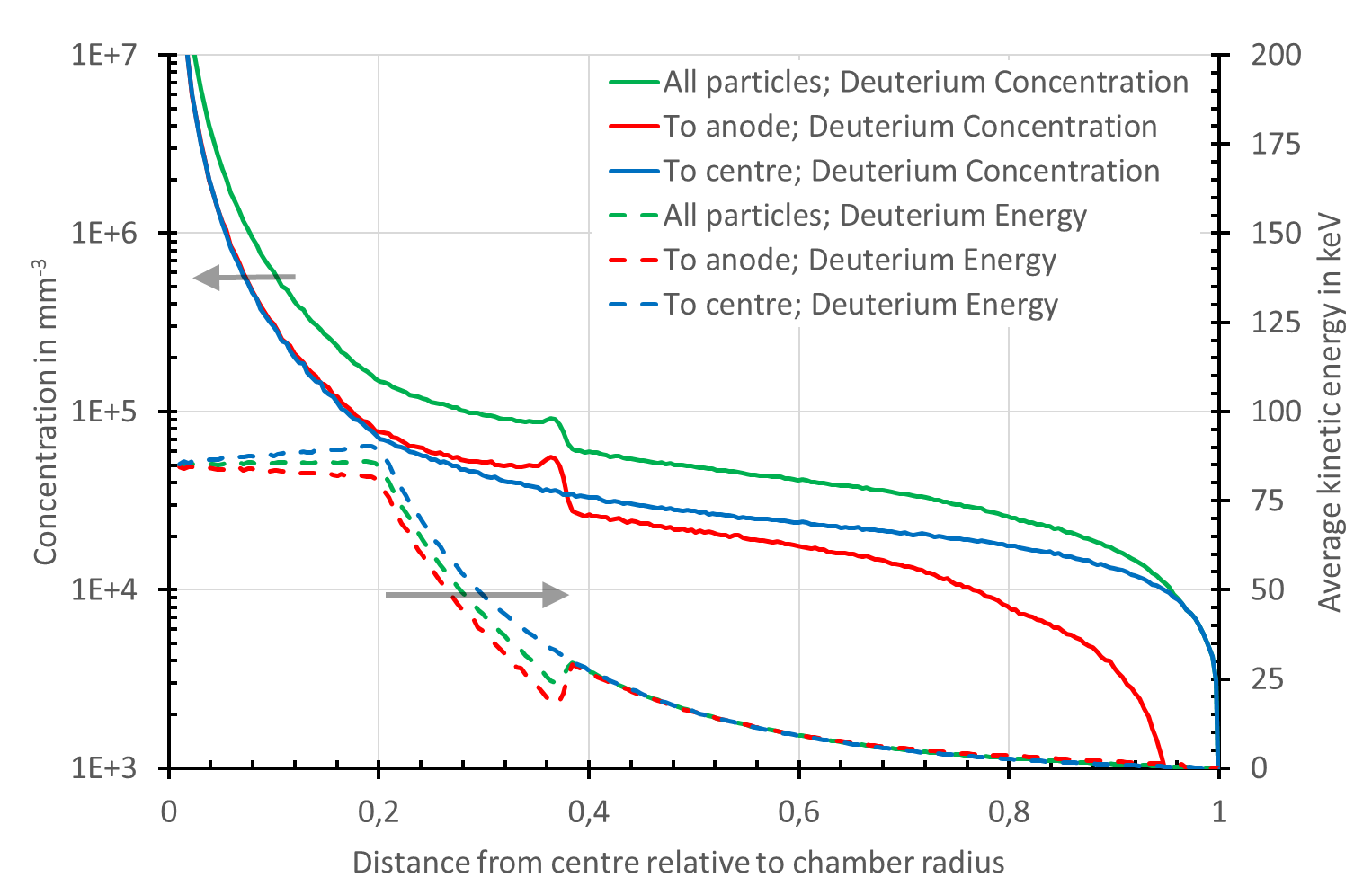}}}
\put(-5,0){\scalebox{0.33}{\includegraphics{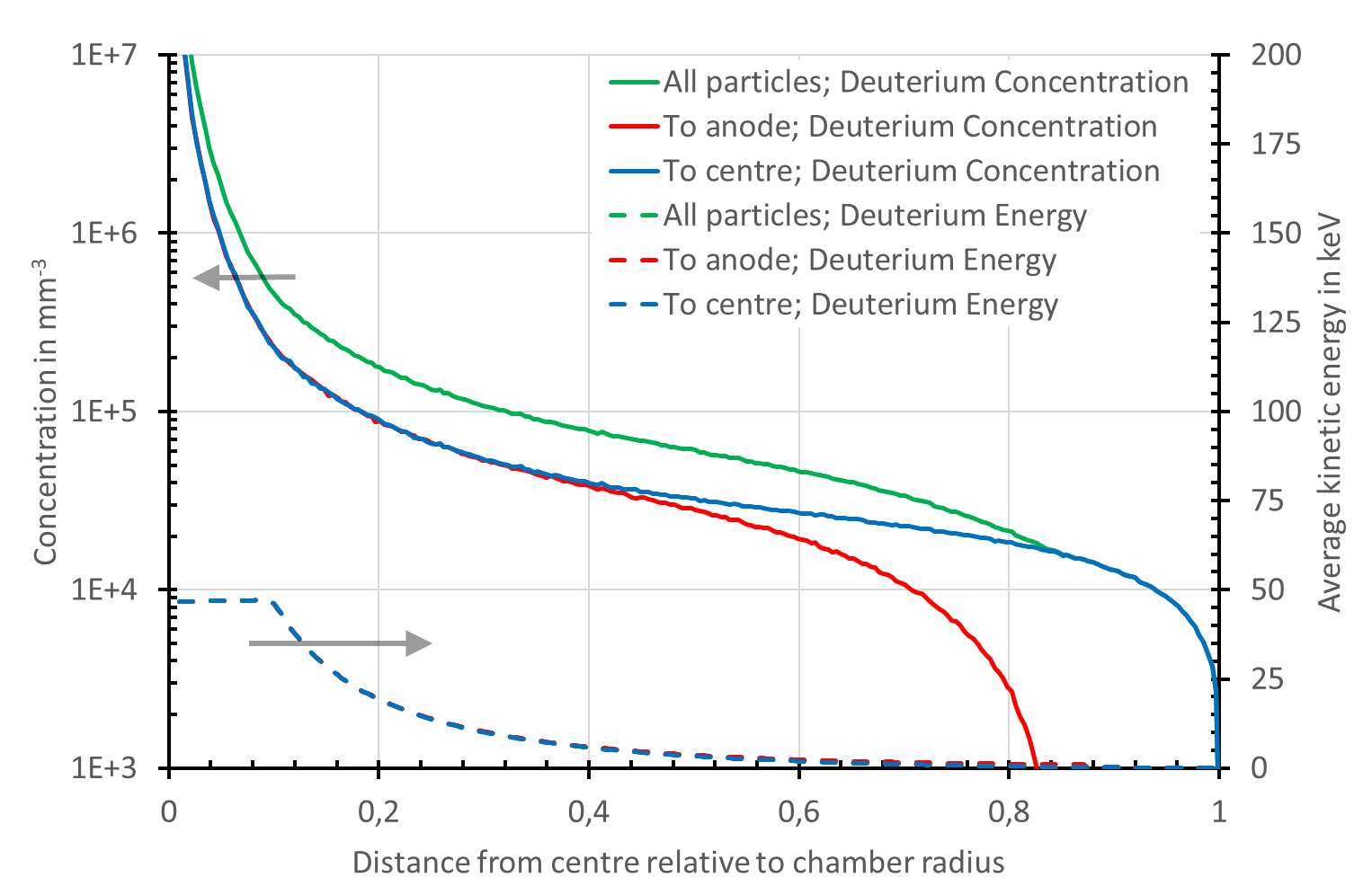}}}
\put(255,0){\scalebox{0.33}{\includegraphics{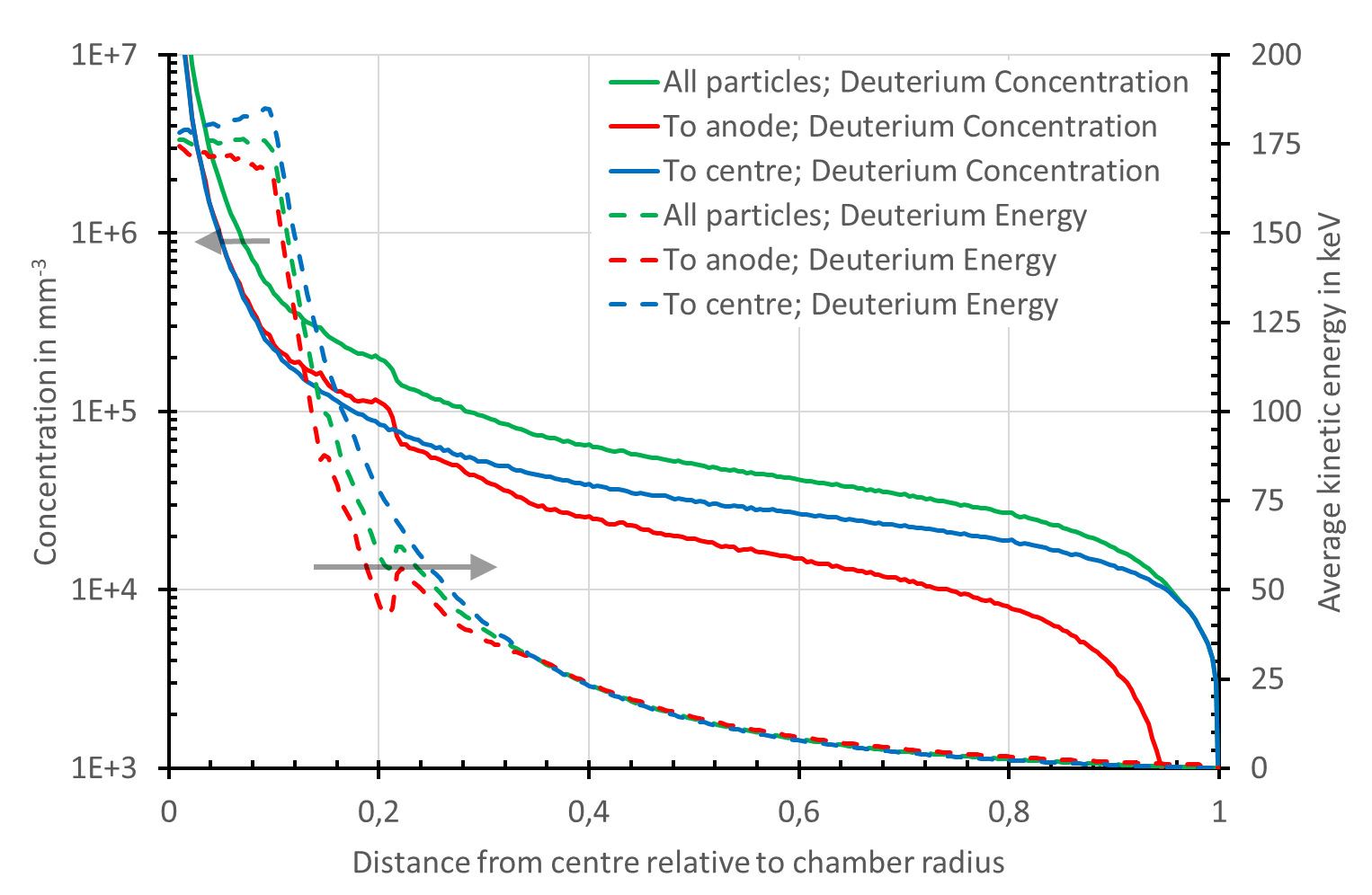}}}
\end{picture}
\caption{\label{figEandCDirection} Deuterium average energy and concentration distribution as function of relative distance to the centre of a fusor with radius of 5 cm for the 4 simulations: Top Left: 5 mm -100 kV; Top Right: 10 mm -100 kV; Bottom Left: 5 mm -50 kV; Bottom Right: 5 mm -200 kV. The left vertical axis shows the concentration (full lines) and the right one shows the average kinetic energy (dashed lines).}
\end{figure*}

Both the deuteron density and concentration exhibit a bump at a location that seems to depend on the radius of the cathode and the applied voltage. When the concentration increases, the average energy decreases, which indicates a slow down of the Deuterium ion. To inspect this further, in figure~\ref{figEandCDirection} the deuteron density and concentration is presented, split into the particles that are moving towards the centre (accelerating Deuterium ions) or away from the centre (towards the anode: decelerating deuterium ions).

These figures expose that the concentration and average kinetic energy of the ions moving away from the centre is different from those moving towards the centre, reflecting the dynamics of these particles. Ions that move towards the centre are accelerated and gain energy. Ions that move away from the centre loose energy until they have such a small velocity that the chances that they recombine with an electron increase dramatically and they are converted to molecules again and are lost for the ion concentration. This is why only the concentration of the Deuterium ions moving away from the centre have a bump: when they decelerate their concentration increases, but when they stop they are removed and the concentration reduces again. 

For the simulation with a voltage of -50 kV their is no clear bump, the ions can almost reach the anode again, but not completely. This means that the ions still have some energy and/or momentum exchange with the molecules. This is also evident in the other simulations. The bump in those simulations show however that the energy and/or momentum exchange of the ions with the molecules is larger when the energy of the ions is larger, reflecting the typical behaviour of the elastic and inelastic cross sections. 

The different slopes of the average energy of the inward and outward moving ions inside the volume of the cathode is also an indication of the average energy loss of the ions while moving through the background gas, that is obviously larger when the energy of the ions increases.

\subsection{Interaction mechanisms}
From the above results and especially from the results presented in figure~\ref{figEandCDirection} it is clear that interaction mechanisms play an important role in the concentration and kinetic energy distribution of the particles and hence of the fusion yield. The influence of the recombination of ions with electrons has been exposed as yielding a more complicated structure of these distributions. Hence, it is important for an accurate simulation to take these issues into account. 

It also possible that one uses the information obtained from these concentration and kinetic energy distributions to optimize yields of other fusion reactions of for instance beam-target systems. By placing electrodes at specific locations inside the fusor so that the average energy of the Deuterium ion reaches an optimal value for the fusion reaction, the fusion rate may be enhanced~\cite{Kilic2026}. 

One does not have to focus on a single fusion reaction but might add additional interaction mechanism, an example of such an mechanism is given by \cite{Czerski2024, Dubey2025}. When a suitable solid state target with a high density of Deuterium is placed at a location where the Deuterium energy is close to the resonance energy one might expect a high sensitivity for this reaction.

\section{Conclusions and Recommendations for Future Developments}
The presented simulations calculate fusion yields similar to those found in real world fusors. In addition, the ion energy and density can be simulated. Given that these ion properties arise from a full Maxwell solver this part of the simulation is accurate given the restrictions based on the introduced simplifications including not-modelled grid losses, simplified ionization/re-injection, and the classical treatment. Hence, the agreement with experimental fusors is qualitative and further work to establish the accuracy is needed. Quantitative comparison with an identified fusor measurement under comparable voltage, pressure, geometry, and discharge current is needed to obtain the full extend of the possibilities of this kind of simulations. The influence of the imposed ionisation fraction and the omitted cathode interception are some of the limitations that needs further scrutiny.

The fusion rate calculation is based on the phenomenological Bosch-Hale formulation that yields cross sections of fusion rates of ionised gases. The simulation presented solves the charge and currents densities of an anode/cathode geometry given a given potential and number of charges given certain conditions and assumptions. The core of the simulations is the 3D Maxwell solver with tracked trace particles. It can be applied for any length scale, provided that the conditions and assumptions on those length scales are realistic. In the current simulation one could also introduce materials with different phenomenological cross sections at certain locations and calculate what the fusion rate would be in a similar way as discussed. 

On the very small length scales the trace particles might be replaced by a quantum mechanical probabilistic interpretation, hence the method described here may also be applied to more complex potentials such as those found in crystals and on surfaces~\cite{Kilic2026}. This can be helpful to determine the total potential produced by the fixed potential provided by the crystal and the component of the potential produced by the mobile ions. In addition, clustering of ions in close vicinity may be observed. Finally, since electrons are also simulated the impact of mobile electrons on clustering and the potential can also be determined. Though, the effects of “heavy” electrons will probably not be seen, since the 3D Maxwell solver is classical and these effects are usually quantum.

Increased fusion yield requires either extreme densities and pressures, which seem unlikely in crystals (nonetheless given the readiness of the simulator that could be looked at), or an amplified cross section which may result from the local potential. As a result, a second simulator which numerically calculates the fusion cross section in a given potential should be constructed. This is necessary to determine whether the charge/plasma structures we observed in the simulated fusor could lead to higher fusion rates. One would have to plug the potential produced by the particles in this simulation into a cross-section simulator to find out if the cross section is significantly modified.

On the experimental side it is possible, for minimal investment, to construct a low energy ($<10$~kV) beam/plasma setup to test low energy fusion enhancements in beam on target and plasma experiments in the vicinity of a Deuterium loaded target material such as Pd or Zr. This circumvents long simulation times and the potential of missing relevant quantities in the simulation. If enhancements are verified it can give a parameter space to sweep through with the simulations.

\section*{Acknowledgement}
\addcontentsline{toc}{section}{Acknowledgement}
The author wants to thank Niels Geerits who contributed to the work performed while under contract of BonPhysics Research and Investigations B.V.
\section*{Author Contributions}
All authors have read and agreed to the published version of the manuscript.
\section*{Conflicts of Interest}
The authors declare no conflicts of interest regarding the publication of this paper.

\bibliography{main}

\end{document}